# A Novel Real-Time Full-Color 3D Holographic (Diffractive) Video Capture, Processing And Transmission Pipeline Using Off-The-Shelf Hardware

*Ankur Samanta\*, Gregor Mackenzie\*, Tyler Rathkamp\*, Adrian Cable\*, Darran Milne\*\*,
Andrzej Kaczorowski\*\*, Ronjon Nag\**

**\*R42 Institute, Palo Alto, CA, USA     \*\*VividQ, Cambridge, UK**

## Abstract

This paper details the world's first live 3D holographic (diffractive) video call using off-the-shelf hardware. We introduce a novel pipeline that facilitates the capture, processing, and transmission of RGBZ data, using an iPhone for image and depth capture with VividQ's SDK for hologram generation and hardware for display.

## Objective

The human visual system uses four cues to perceive depth in a scene: accommodation, convergence, binocular parallax (stereopsis), and monocular parallax. The vast majority of self-proclaimed "three-dimensional displays" rely on the notion that not all of these cues need to be present to convey depth. However, when some of these cues are missing, viewer discomfort (including eye strain and nausea) usually results.

Holographic displays provide the only practical way to present three-dimensional content using all four of these cues. However, the current suite of attempts at generating 3D augmented/virtual reality depictions of real-world objects rarely utilize true holographic projection, because of optical system complexities and the constraints of real-time computation of holograms from 3D input data. In recent years, however, there has been significant progress in this field, with "ready for market" near-to-eye displays and other true holographic displays emerging from suppliers including VividQ.

Much research has already been undertaken on the side of computer-generated graphics and visual representations of modeled 3D data, with companies generally creating their own specialized approach and equipment for generating 3D depictions of data. However, there has yet to be a generalizable and truly universal 3D data handling pipeline that is mobile, handles depth data efficiently, and can operate in real-time. As such, there is a definite need for the development of such a pipeline to allow for the introduction of true 3D data-related functionality in widespread applications, whether in telecommunications, modeling, AR representations, data storage, etc.

When it comes to visualizing the impact and importance of real 3D data handling capabilities for display technology, real-time conferencing-type applications are both efficient and compelling as a test case. Many groups have tried to emulate the full visual impact of 3D hologram streams, but almost all of them use non-holographic techniques like the Pepper's Ghost illusion or stereoscopic display viewing. These approaches suffer from numerous shortcomings, like not accommodating high-accuracy depth perception using all four visual cues, something that is essential to true 3D data display and manipulation applications. Thus, as a proof of concept for the implementation and live capability of such a 3D data handling pipeline, this paper details the implementation of the world's first live 3D diffractive holographic video call, as facilitated with the novel 3CPT pipeline (3D Capture, Processing, Transmission).

## Background

There have been a number of technological limitations that have hindered the development of such a pipeline until now, in the context of 3D hologram generation.

One key issue with the uptake of holographic displays is that generation, processing, and transmission of video data in three dimensions necessarily involves a fundamentally different approach than two-dimensional image processing. Most conventional video codecs on the market (used for compression and decompression of video streams) do not support 3D data, making the widespread usage of 3D data in remote imaging applications that much more difficult. As such, there is a need to develop approaches to 'piggyback' three-dimensional image capture, processing, and transmission on existing architectures that have been designed for 2D video.

Additionally, most setups that allow for depth data capture rely on expensive and highly specialized 3D camera equipment, a cumbersome prospect when considering the demands of scalability for the mass market. This notion deserves re-examination following the more recent mass introduction of depth-capable cameras in smartphones, for example, the TrueDepth camera introduced in the Apple iPhone X. This dramatic increase in the accessibility of depth-capture-capable devices provides for a unique opportunity to make high-accuracy 3D data capture and processing possible from almost anyone's pocket.

And finally, another key technological constraint is the display medium. VividQ has made it possible to display true 3D images using computer-generated holography, which serves as a basis of development on which real-time image capabilities can be built. We now endeavor to develop an end-to-end pipeline that can connect a smartphone from anywhere in the world to a 3D hologram projector and as such enable the live-streaming of 3D image data captured in real-time.

## Methods

### Architectural Overview

We propose a novel pipeline, 3CPT, for the real-time capture, processing, and transmission of true 3D data, that can be used to power any application that requires depth-aware, low-latency, and high-resolution 3D data handling in real-time. The pipeline can be linked to any image capture platform but is designed to be able to generate, encode, transmit, and decode high-quality 3D data streams from any smartphone with depth-capture capabilities (i.e a dedicated depth camera or LiDAR sensor). This allows for relatively low-cost, modular, and mobile 3D imaging from the capture end.

3CPT is primarily designed to support the generation, transmission and replay of real 3D holograms, having been recently used to connect smartphones around the world to 3D hologram projectors based on VividQ's platform. This feat demonstrated the viability of live 3D holographic conferencing as an emerging marketable technology in the near future, but the pipeline itself is developed in a way that allows it be retooled and deployed for almost any application relying on 3D data - including ones more market-ready in the more immediate future.

Following is a breakdown of the implementation of 3CPT as a standalone pipeline, after which its deployment and proof of viability in facilitating the world's first live 3D holographic video call will be detailed.

*Implementation*

A fully integrated 3CPT pipeline can be divided into 4 segments: Capture, Processing, Transmission, and Reception. These segments are then implemented across 3 terminals: Capture, Intermediary, and Reception, which is connected to the relevant display medium.

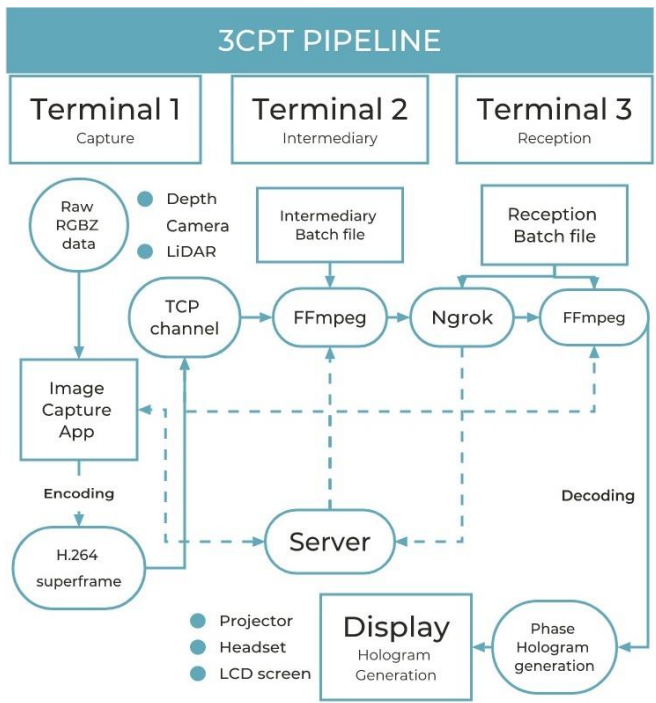

**Figure 1.** 3CPT High-level Schematic

For data capture, the user installs the 3CPT app, which taps into the depth+main cameras on a mobile device, enabling the capture of full 3D data (color and depth) - for greater detail, there is a LiDAR option in development, but standard depth cameras on smartphones offer greater cross-compatibility in the mean-time. For comparison, other setups involving 3D cameras might have multiple 1080p video cameras and an IR laser scanner for depth, which is far less scalable/feasible for mass use [1]; however, note that the pipeline can easily be repurposed to also work with more complex 3D imaging setups for higher-precision live data modeling, should it be deemed necessary. Moving on, the 3CPT app then utilizes the depth-data extraction functionality present in the iOS AVFoundation SDK [2] to extract disparity data from the source stream, use orientation data to align it, use filtering to fill in the gaps and smoothen/sharpen parts of the image, normalize the depth data, and return it as a separate depth map image.

For processing, the app then filters out background noise and reformats the data (still in standard video format) into a 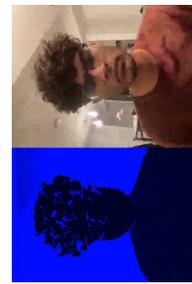 "superframe" featuring two 2D frames ("depth map" image stacked underneath a regular "brightness/color" image). This enables synchrony between the video and depth without having to simultaneously reconstruct and concatenate multiple streams. This single stream transmission approach means that depth data can be compressed using just RGB, rather than needing a custom RGBZ (RGB+depth) process for compression, encoding, and decoding.

**Figure 2.** Stacked Frame

Conventional video codecs are designed for RGB/YUV 2D input and are not able to handle 3D input, presenting a core challenge for the implementation of such a pipeline. Now, instead of resorting to developing a new 3D-capable codec from scratch, time and efficiency considerations by the team led to the evaluation of conventional codecs operating on "superframes" as described above. For this rendition of the pipeline, H.264 was used due to its ubiquity and access to an effective reference encoder (libx264), although other codecs such as VP8/9 and H.265 would also be well suited.

The transmission component utilizes a combination of FFmpeg and ngrok, a cloud relay library for creating TCP connections between two network endpoints that would likely otherwise not permit direct connectivity, due to NAT, firewall, or other limitations. A separate cloud server is also used to facilitate communication of relevant port keys and IDs between the three terminals of the pipeline, allowing for automatic instantiation of connections and enabling multiple simultaneous streams under different channel IDs. Wrapping up the capture end, the 3CPT app outputs the newly formatted 3D data to the specified TCP port used by ngrok.

On the receiving end of the pipeline, the listener receives the data from the ngrok transmission and decodes the data from H.264 to a standardized format that can be interpreted and converted into a phase hologram, while maintaining the brightness, color, and depth aspects of the video data. This is then fed to the hologram generation pipeline, which in this case is an API from VividQ's Software Development Kit (SDK) for hologram generation.

The VividQ SDK calculates computer-generated holograms out of the color-plus-depth data in real-time (a highly computationally intensive process that runs on NVIDIA GPUs). The holograms are then sent to a phase-only Spatial Light Modulator (SLM), which is illuminated by RGB lasers to form a true 3D image with depth perception in space. This 3D hologram can be viewed through an optical assembly, which can be tailored to the exact application the pipeline is being used to facilitate. Clients have a choice of using custom-tooled holographic projectors, holographic AR headsets, or holography-enabled LCD screen displays (with an emphasis on cross-compatibility with different display mediums to accommodate the inevitable growth of the holographic display space).

As for customization, the application dictates the parameters and display medium setup. In the case of holographic video conferencing, the emphasis was on having the user be able to speak to a hologram as if they were conversing with it. The available technology meant that a holographic projector made the most sense (using largely off-the-shelf parts), which was configured in partnership with VividQ to simulate the experience of a traditional projector as well as a headset.

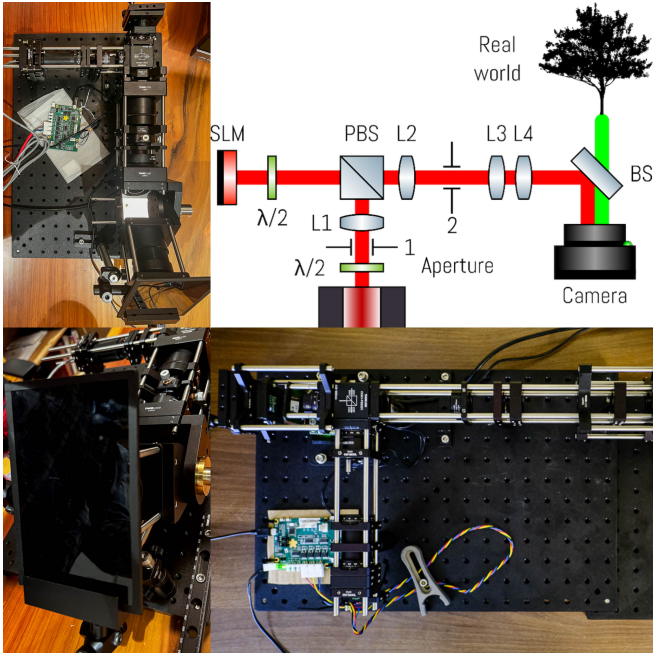

**Figure 3.** (left, top/bottom): demonstrates a traditional projector setup, where the user looks into a filtered viewing pane to observe the 3D hologram. **Figure 4.** (right, top/bottom): demonstrates a projector setup where the camera setup is done so as to emulate the experience of a holographic headset, where the hologram appears to float in the surrounding environment itself.

*Deployment*

The 3CPT pipeline was used to demonstrate the world's first live 3D holographic video call in real-time. In conjunction with VividQ's hologram generation technology, the pipeline was instrumental in developing a modular, mobile setup that allowed a user with a depth-capable smartphone to connect to a hologram projector from anywhere in the world, and have their real-time, 3D likeness projected in front of the user on the other end. Note that this pipeline can be retooled for and deployed in almost any application that requires high-quality, low-latency, live 3D data streaming/reconstruction, with cross-platform compatibility with different capture and hologram generation mediums.

*Technical Specification*

When tooling the pipeline for specific applications, different data parameters need to be considered, including field-of-view, resolution, packet size, latency, and frame rate.

As the pipeline currently stands, the iPhone camera outputs video at a 640x480 pixel format and so this is also the format of RGB+depth data received by the endpoint after decoding. The VividQ SDK is designed to compute holograms for replay on a 2048x2048 SLM, but due to the limitations of binary holograms a maximum effective field size of 2048x1024 is available. As such, for simplicity, we scale the 640x480 image up to 1280x960, and then embed it in a 2048x1024 array of zeros, before zero-padding to 2048x2048 elements and passing to the VividQ SDK, with each element consisting of 4 bytes: R, G, B, and Z (a disparity value in diopters, which is calculated from the iPhone camera's depth data).

This technique could be improved by using bilinear or cubic resampling for the scale operation, although even as currently implemented it results in a clear image without having to distort the video from our iPhone camera. For further improvements in quality, built-in smartphone LiDAR sensors can be used (requiring a different encoding stack, but feasible nonetheless), as well as more complex 3D imaging setups as required by the application.

**Results**

This pipeline was used to power a live 3D holographic stream, with image capture done using an iPhone and the 3CPT app from Toronto and Palo Alto, and hologram generation done with a VividQ hologram projector in Cambridge.

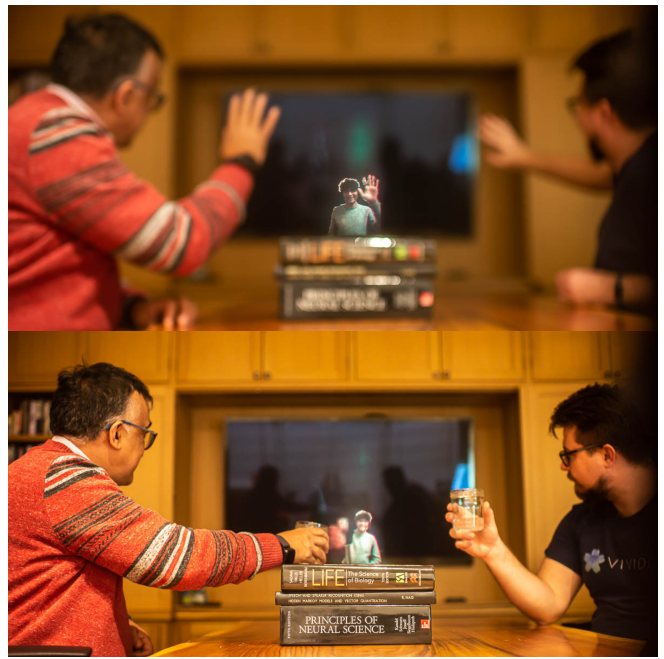

**Figure 5.** (top): Hologram waving; (bottom): Holographic "Cheers" demonstrating perceived interaction with real-life subjects - varying degrees of focus achieved due to 3D nature

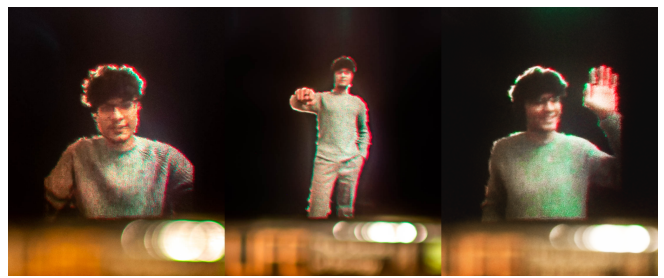

**Figure 6.** Series of holograms demonstrating different poses and motions at different depths

**Figure 7.** (top): Camera lens focused on different parts of the image at different depth levels - the difference in clarity is achieved relative to the plane of depth, as is the case with real-life 3D objects. (bottom): Multiple subjects

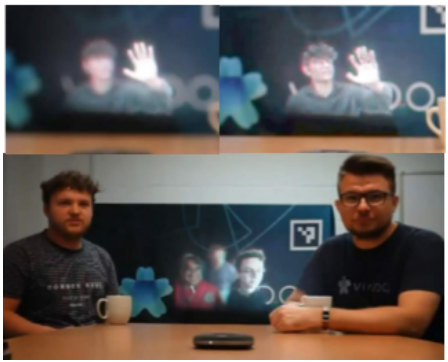

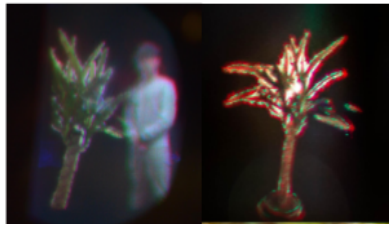

**Figure 8.** (left): Looking into the projector to see a floating hologram appearing out of the viewing window. (right): Hologram appears to float on existing objects in the surroundings (akin to a holographic headset)

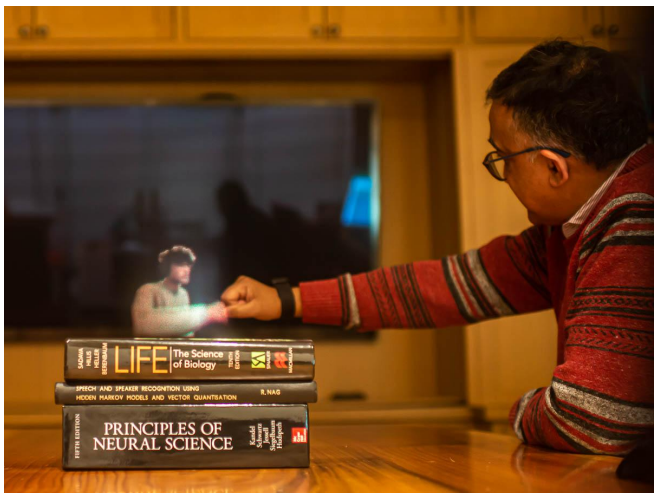

**Figure 9.** Holographic fistbump, showcasing perceived human-hologram interaction, visualizing the vision for future integration of holographic technology into our day-to-day lives

Moving on to the technical performance, while testing, we noticed a bottleneck in the pipeline experienced due to the sheer amount of data being sent over local networks in the initial step. For reference, each frame consumed from the iPhone camera takes 2457600 bytes (640*480*2*4). The internal image format for the frame is RGB0 (4 bytes per pixel; the last byte is redundant) to achieve optimal alignment to form the superframe. These may be split between multiple TCP packets due to limited MTU, so frame reassembly from fragmented packets was also handled at that point.

We found during testing that the principal performance bottleneck in the entire system was the transmission of data over the local network between the iPhone used for depth capture and the Mac used for transcoding, rather than the transcoding itself.

This led to some initial lag during the initialization of the video stream, which resolved itself after a nominal period of time, after which near-perfect synchronization with the live feed was achieved, proving the real-time capabilities of the pipeline. This was proven via a demonstration of subsecond delay between the visual of a clap transmitted holographically and its sound sent through an independent low-latency channel e.g. Zoom.

In addition, the hologram generated using live capture was tested against similar computer-generated holograms to contrast the rendering quality between the two images, and the live image quality was largely comparable. For video demonstration (using the setup from figure 3) of the aforementioned tests and other visuals, please scan the QR code to the right.

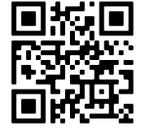

Overall, multiple live streams have since been conducted successfully and while further development will no doubt result in an even lower-latency, higher quality holographic livestream, this demonstration proved the viability of the 3CPT pipeline as a generalizable tool to facilitate 3D data functionality in a wide variety of applications, even highly demanding ones such as videoconferencing.

## Impact

The advent of true 3D holographic technology has the potential to have a massive impact on the display technology industry, and our lives in general. Instead of creating artificial realities to view through headsets and trying to mimic real-life interactions in a virtual domain, this technology allows us to integrate real 3D holographic representations in the real world. This is true augmentation of the real-world experience, which is far better than trying to recreate that experience in a virtual world that detaches us from reality. Additionally, the possibilities of facilitating human-hologram interaction with third-party tools such as haptics will make the experience even more realistic, with the overall emphasis being on the augmentation of our real lives and world.

As for the 3CPT pipeline, this is the first time that such technology has been facilitated using an easily accessible device like a smartphone. Most attempts at "holography" either aren't actual holograms and therefore are not able to present all four depth cues, or are not easily scalable/replicable for mass usage. A pipeline that can use an iPhone, or any other depth-capable smartphone, to provide it with the necessary data instantly opens up both a wide range of possibilities and applications, as well as flexibility in the business model being used.

3D holographic live streaming can make 3D video conferencing a real possibility, marking a revolutionary shift in the teleconferencing industry, aside from the large number of other industries where 3D data handling capabilities could usher in the next stage of development and evolution of display technology.

## Prior Publications
None relevant.